\documentclass[prd, reprint, superscriptaddress, preprintnumbers, nofootinbib, amsmath, amssymb, aps]{revtex4-1}
\pdfoutput=1
\usepackage{hyperref}
\usepackage{graphicx}
\usepackage{makecell}
\usepackage{subfigure}
\usepackage[textwidth=1.25cm,textsize=footnotesize]{todonotes}
\usepackage[utf8]{inputenc}
\usepackage{color}
\usepackage[normalem]{ulem}

\begin{document}
\title{Search for a photon peak from keV-scale dark matter annihilation with NuSTAR: Constraints on $\langle \sigma v \rangle$ after 11 years of observations}

\author{E.I. Zakharov}
\email[]{ezakharov@cosmos.ru}
\affiliation{Space Research Institute of the Russian Academy of Sciences, Moscow 117997, Russia}
\affiliation{National Research University Higher School of Economics, Moscow 101000, Russia}
  
\author{V.V. Barinov}
\affiliation{Institute for Nuclear Research of the Russian Academy of Sciences, Moscow 117312, Russia}

\author{D.S. Gorbunov}
\affiliation{Institute for Nuclear Research of the Russian Academy of Sciences, Moscow 117312, Russia}
\affiliation{Moscow Institute of Physics and Technology, Dolgoprudny 141700, Russia}

\author{R.A. Krivonos}
\affiliation{Space Research Institute of the Russian Academy of Sciences, Moscow 117997, Russia}

\author{A.A. Mukhin}
\affiliation{Space Research Institute of the Russian Academy of Sciences, Moscow 117997, Russia}
\affiliation{Moscow Institute of Physics and Technology, Dolgoprudny 141700, Russia}

\bigskip
\preprint{INR-TH-2025-013}

\begin{abstract}
We report new constraints on the velocity-independent annihilation cross section $\langle \sigma v \rangle$ of keV-scale dark matter particles based on 11 years of observations with the NuSTAR X-ray telescope. Using the unfocused stray light mode of the instrument, which provides a wide field-of-view and a stable instrumental background, we perform a sensitive search for photon signatures from dark matter annihilation in the Galactic halo. We model the resulting diffuse X-ray spectrum over the 3-20 keV energy range and search for line-like spectral features that may arise from the annihilation of dark matter particles into photons. No statistically significant excess over the expected astrophysical background is found. We therefore place upper limits on $\langle \sigma v \rangle$ as a function of dark matter mass, assuming a velocity-independent s-wave annihilation and several Galactic dark matter profiles. Across most of the explored mass range our results provide the strongest X-ray constraints to date, reaching the level of $\langle\sigma v\rangle \lesssim 10^{-33}$--$10^{-34}$\,cm$^3$\,s$^{-1}$, and they are complementary to the most recent bounds derived from SRG/ART-XC observations.
\end{abstract}

\maketitle

\section{Introduction}

Numerous Dark Matter (DM) phenomena in astrophysics and cosmology call either for a specific gravity modification or new particle physics. The latter variant implies an extension of the Standard Model of particle physics which produces a sufficient amount of non-relativistic, electrically neutral and stable on cosmological time-scale matter in the early Universe well before the temperature of cosmic plasma drops below 1\,eV, see e.g.\,\cite{Rubakov:2017xzr}. The matter may consist of some macroscopic objects, e.g. primordial black holes, $Q$-balls, or just new fundamental particles. The latter case can be generically tested with both direct and indirect searches for these new particles. For a recent comprehensive review on the subject see Ref.\,\cite{Cirelli:2024ssz}.  

A variety of models of DM particles suggest candidates with masses from $10^{-21}$\,eV (fuzzy DM \cite{Hu:2000ke,Bar:2018acw}) to the Planck scale (maximons \cite{Markov:1967lha}). Different models can be tested in different ways, but as far as cosmology and astrophysics are concerned, there are no any clear preferences of the DM mass scale within the broad interval indicated above. In this situation, naturally, the extensive hunt for the DM particles is unceasing and is limited by experimental tools at hand rather than by theoretical prerequisites.  

In this paper we consider the X-ray orbital telescope NuSTAR as the sensitive tool to test the models with DM particles of keV-scale masses (see e.g.\,\cite{Goudelis:2018xqi,DEramo:2020gpr,Cheek:2024fyc}). The Galactic DM particles may annihilate into a couple of photons, 
\begin{equation}
\label{annihilation}
\text{DM}+\text{DM} \to \gamma+\gamma\,,
\end{equation}
that provides with a peak signature in the Galactic X-ray diffuse spectrum. Different models predicting such annihilation are known in the literature, e.g. \cite{Brdar:2017wgy, Dudas:2014ixa}. The peak frequency must be equal to the DM mass, 
\begin{equation}
    \omega_\gamma=m_\text{DM}\,,
\end{equation}
up to the Doppler broadening at the level of $10^{-3}$ due to non-zero velocity dispersion of DM particles in the Galaxy.

Any observation of such a peak may be interpreted as a signature of DM Aannihilation, pins down the DM mass and gives an estimate of the effective DM-DM-$\gamma$-$\gamma$ coupling. Absence of the suggested signature places a corresponding limit on the model parameter space. 

It should be noted that the possible radiative decay of the DM, like the one inherent in the model of sterile neutrino DM\,\cite{Drewes:2016upu}, implies a similar peak-like signature in Galactic X-rays (see Ref.\,\cite{Krivonos:2024yvm} for the most recent investigation). However, the two models, decaying DM and annihilating DM, can be distinguished by tracing the variation of the signal intensity over the sky: the signal concentration in the Galaxy center region is more pronounced in the latter case.

The analysis presented below, based on the same data set and methods from our previous work \,\cite{Krivonos:2024yvm}, yields the most stringent X-ray constraints currently available on keV-scale dark matter annihilation.

\section{The NuSTAR Observatory and Data Set}
\subsection{The NuSTAR Telescope}
The \textit{Nuclear Spectroscopic Telescope Array} (NuSTAR) is a hard X-ray observatory launched by National Aeronautics and Space Administration (NASA) in 2012, operating in the energy range of 3-79 keV \cite{NuSTAR}. Being the first focusing high-energy X-ray telescope in orbit, it carries two co-aligned, independent telescope modules, referred to as FPMA and FPMB. Each module comprises a multilayer-coated grazing incidence Wolter-I optic and a focal plane detector, enabling imaging of the sky with an angular resolution of 18$''$ (FWHM) and a field of view (FoV) of approximately 13$'\times$13$'$.

The focal plane consist of a 2$\times$2 array of CdZnTe (CZT) pixel detectors, each with 32$\times$32 pixel, providing moderate energy resolution ($\sim$400 eV FWHM at 10 keV) and high quantum efficiency across the operational range. However, an important feature of NuSTAR for diffuse background studies is the presence of a stray light (SL) aperture -- an open geometry between the optics and the detectors that allows unfocused photons to directly reach the detectors from directions several degrees off-axis (typically 1$^\circ$--3$^\circ$)\cite{2017JATIS...3d4003M}.

Although originally considered a contaminant in pointed observations, the SL component has proven to be a valuable probe of diffuse X-ray backgrounds. It effectively transforms the telescope into a wide-field spectrometer. For our purposes, the SL mode provides access to a large solid angle on the sky with a stable and well-characterized instrumental response. Combined with the long operational baseline of NuSTAR, this enables the accumulation of high photon statistics, making the SL signal particularly suitable for indirect DM searches based on spectral features.

An important figure of merit for observations targeting large angular scales is the grasp, defined as the product of the effective detector area and the average solid angle subtended by the FoV. In the SL configuration, the effective area is determined purely by the geometrical size of the focal plane detector, which is approximately 13 cm$^2$ per module. The average solid angle, $\Omega_\text{SL}$, over which the detector is exposed to SL photons, is about 4.5 deg$^2$. This yields a nominal grasp of $G = \Omega_\mathrm{SL} \times A_\mathrm{SL} \approx 58$ deg$^2$ cm$^2$ for each of the FPMA and FPMB modules. However, data processing steps, such as the bad pixel removal the source masking, reduce both the usable detector area and the effective solid angle. For typical clean observations, the effective area decreases to approximately 10 cm$^2$, while the solid angle is reduced to about 4 deg$^2$, resulting in an effective grasp of $G \approx 40$ deg$^2$ cm$^2$ per module \cite{Perez_2019}. Despite this reduction, the grasp in stray light mode remains substantially larger than the $\sim$8 deg$^2$ cm$^2$ grasp associated with photons focused by NuSTAR mirror system \cite{Roach:2022lgo}. This enhanced sensitivity to diffuse flux makes the stray light aperture a valuable tool for studies of extended X-ray emission, including indirect searches for DM signatures on degree angular scales.

\subsection{Observation Campaign and Data Selection}
To construct a high-exposure dataset suitable for the analysis of diffuse X-ray emission in the SL mode, we utilized the archival data from the NuSTAR mission spanning the period from July 2012 to January 2024. The initial selection included all publicly available observations, with the exception of those specifically targeted at Solar System objects or characterized by insufficient exposure time.

In particular, we excluded observations with ObsIDs starting with the digit 2, which correspond to dedicated observations of the Sun and other nearby sources, as well as two Jupiter-related datasets acquired under director’s discretionary time (ObsIDs 90311 and 90313). Furthermore, we removed all exposures with a total duration below 1 ks, as such short observations contribute little to the overall photon statistics and are more susceptible to background uncertainties.

The resulting filtered set was further processed to remove instrumental artifacts and residual focused emission using an automated wavelet-based masking procedure. For more details see section \ref{remove}.

\begin{figure}
    \centering
    \includegraphics[width=1\linewidth]{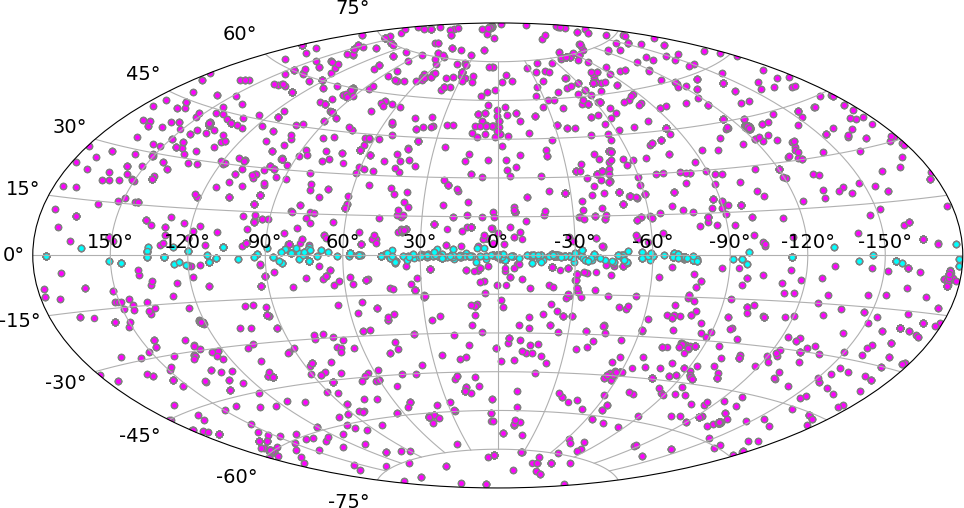}
    \caption{The distribution of 3248 (FPMA) and 3139 (FPMB) NuSTAR observations on the sky in Galactic coordinates. Cyan and magenta points show the NuSTAR observations at $|b|<3^{\circ}$ and $|b|>3^{\circ}$, respectively.}
    \label{fig:sky_dist}
\end{figure}

\begin{figure}
    \centering
    \includegraphics[width=1\linewidth]{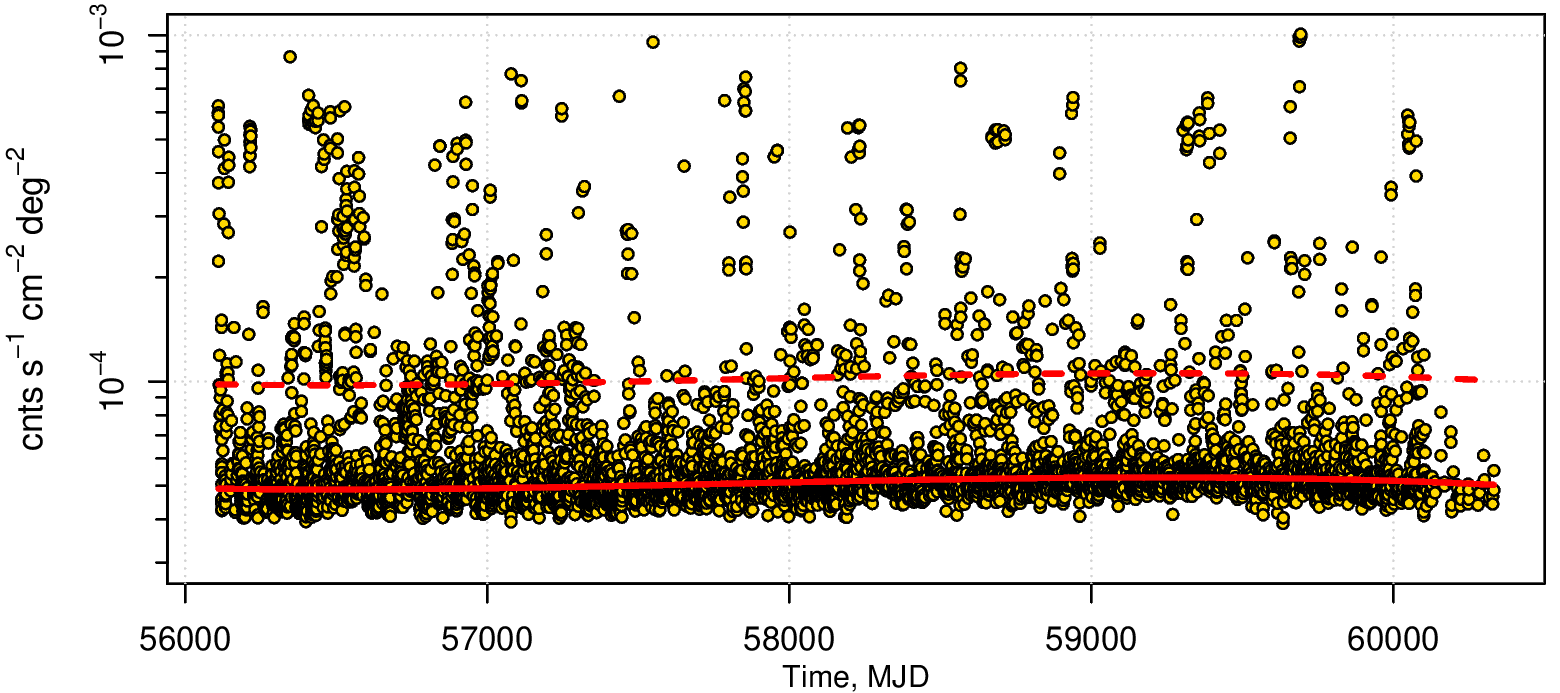}
    \caption{The NuSTAR detector count rate as a function of time, MJD stands for Modified Julian Date. Each point represents individual NuSTAR observation. A solid red line is a cubic polynomial outlier-resistant approximation used to describe long-term variation. A dashed red line shows the same approximation scaled by an arbitrary-chosen factor of 2, that is a threshold: the observations above it are excluded from the analysis, since the high count rate is due to the remaining contribution of the GRXE at slightly higher Galactic latitudes than $|b| \sim 3$ deg.}
    \label{fig:variability}
\end{figure}

The resulting list is comprised of 3248 observations for FPMA (with a total exposure of 150.1 Ms) and 3139 for FPMB (144.9 Ms). To minimize contamination from the Galactic ridge X-ray emission (GRXE), we exclude all observations with Galactic latitude $|b| \leq 3^\circ$. The distribution of observations across the sky is shown in Fig.\,\ref{fig:sky_dist}. A basic temporal filtering with arbitrary-chosen threshold was also applied to remove datasets exhibiting clear deviations in the detector background rate from the long-term average (see Fig. \ref{fig:variability}), which helps to avoid remaining contribution of GRXE at slightly higher galactic latitudes than $|b| \sim 3$ deg.

After all filtering steps, we retained a total of 5216 observations (combined FPMA and FPMB), with cumulative usable exposure amounting to 234 Ms. This dataset provides a nearly all-sky coverage at intermediate to high Galactic latitudes and forms the basis for the spectral analysis presented in the following sections.

\subsection{Removal of focused X-rays}
\label{remove}
\begin{figure}
    \centering
    \includegraphics[width=1\linewidth]{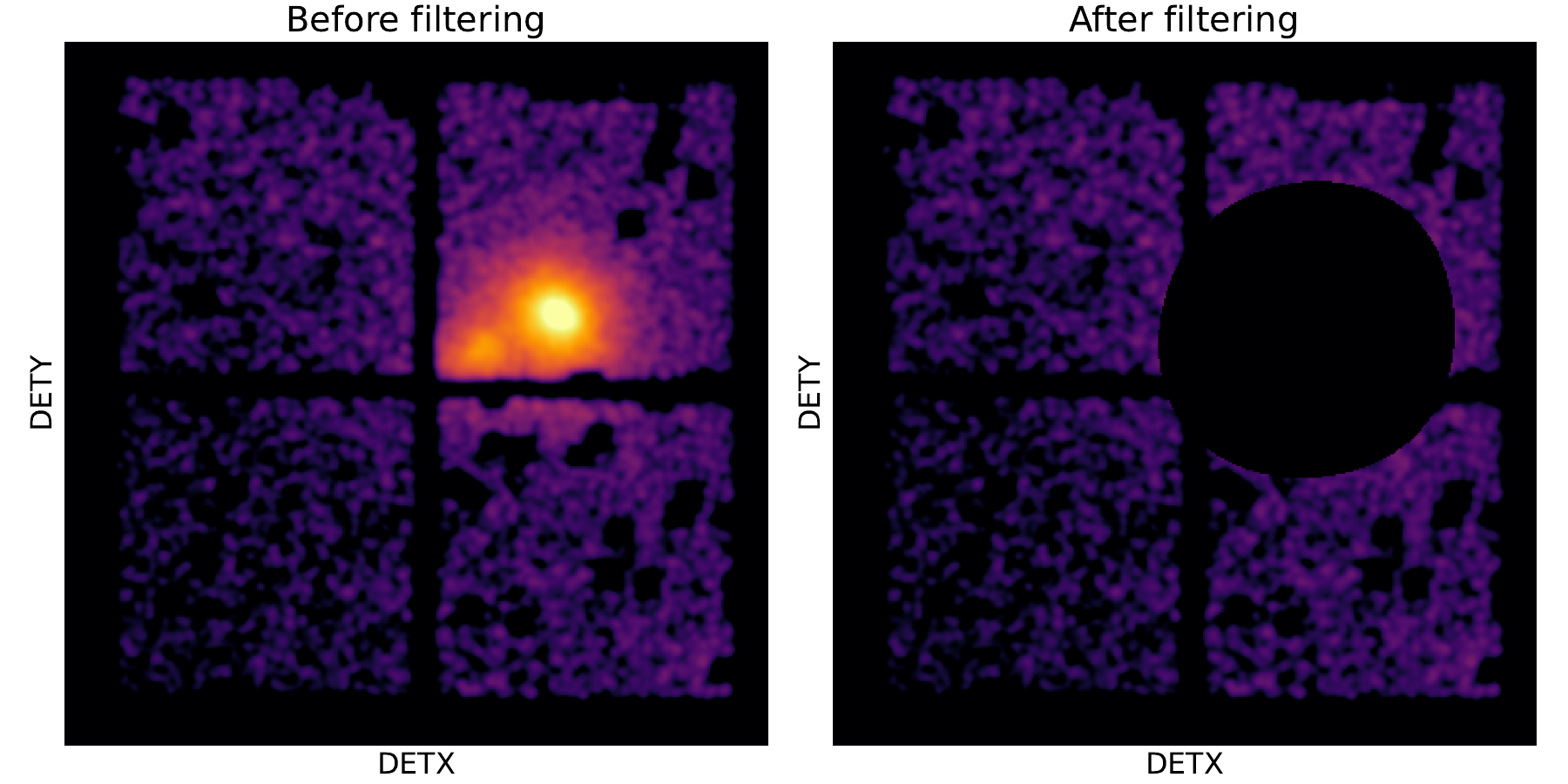}
    \caption{Demonstration of the performance of a wavelet-based algorithm for the removal of focused X-rays. The left panel shows an image of the FPMB detector before the focused X-ray removal procedure was applied. The right panel shows the detector after. The right panel clearly shows the characteristic pattern from the SL (see right panel of Fig. \ref{fig:solidAngle}).}
    \label{fig:wavletDemo}
\end{figure}

A critical step in the preparation of SL data is the suppression of photons that have been focused by the NuSTAR optics. Even though these events represent only a fraction of the total counts, they appear as localized features on the detector plane and, if not removed, can distort the diffuse spectrum and bias the search for faint line-like signatures. To address this, we followed the procedure developed in Refs.~\cite{Krivonos:2020qvl,Mukhin2023}, where an automated wavelet-based algorithm is used to identify and mask compact structures associated with focused point sources. The algorithm effectively separates localized excesses from the smoothly varying SL signal and thus provides a robust way to clean the dataset. Fig. \ref{fig:wavletDemo} shows the detector image before and after applying the focused X-ray removal algorithm.

The quality of each observation after masking was quantified using the modified Cash statistic
\begin{equation}
    C = \frac{2}{N_{\text{bin}}}\sum_i E_i-C_i+C_i \cdot (\log C_i - \log E_i) 
\end{equation}
where $C_i$ is the number of counts for each bin (here we combine the detector pixels into bins until there are at least 2 counts in each bin); $E_i$ is the expected value for each bin, calculated as the expected value for a single detector pixel $I_\text{mean}$ multiplied by the number of pixels in the bin; $N_{\text{bin}}$ is the total number of bins. Observations were retained only if the condition $C < 1.4$ was satisfied, ensuring that the residual focused emission was statistically consistent with the expected detector background.

In addition, we required that at least 40\% of the detector area remained usable after the masking step. Datasets not meeting this criterion were excluded from further analysis. These requirements guarantee both high photon statistics and minimal systematic contamination. After filtering, the retained events are dominated by photons arriving in the SL geometry, while the contribution of residual focused emission is negligible for the purposes of our study.

\subsection{Stray Light spectrum}
\begin{figure}
    \centering
    \includegraphics[width=1\linewidth]{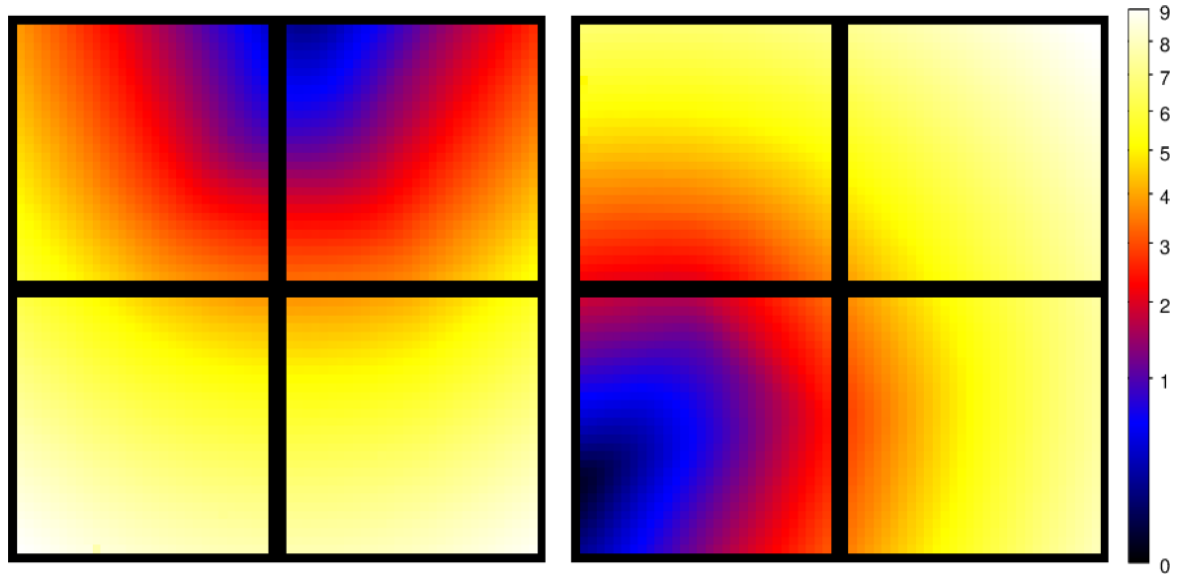}
    \caption{Image of the NuSTAR FPMA (left) and FPMB (right) in physical detector pixels, showing the open portion of the sky for each detector pixel in squared degrees. The image is taken from \cite{Krivonos:2020qvl}.}
    \label{fig:solidAngle}
\end{figure}

After removing the focused X-rays, we are left with only two components on the detector. The first component is a spatially flat detector background. The count rate at a given pixel is proportional to the nearly flat instrumental background, which includes internal emission lines and continuum. The second component is spatially variable aperture background. The count rate of this component is proportional to the open sky solid angle. We can write the number of photons $N$ with energy $E_\gamma$ registered at each $i$-th pixel of the detector during time $T$ as:
\begin{equation}
    N_{\text{pix},i}(E_\gamma) = (C_\text{int} M_\text{int} Q + C_\text{apt} R_\text{pix} \mathcal{E}_\text{tot}A\Omega)_i T.
\end{equation}
Here $C_{\text{int},i}$ is the internal background rate, $M_{\text{int},i}$ describes detector uniformity (see \cite{Krivonos:2020qvl} for details), $Q$ is the relative time-dependent correction factor tracing the long-term radiation environment variation (see Fig. \ref{fig:variability}), $C_{\text{apt},i}$ is the aperture flux per solid angle, $R_\text{pix}$ is the pixel response matrix stored in the NuSTAR CALDB, $\mathcal{E}_\text{tot} = \mathcal{E}_\text{det}\mathcal{E}_\text{Be}$ is the energy-dependent efficiency of the inactive detector surface layer and beryllium entrance window, $A$ is the area of each detector pixel (0.36 mm$^2$), $\Omega_i$ is the open sky solid angle as seen by each pixel in deg$^2$ (see Fig. \ref{fig:solidAngle}), $T$ is exposure time. To construct the spectrum of $C_\text{apt}$ we defined 100 energy bands $E_\gamma$ logarithmically spaced between 3 and 20 keV. 

\begin{figure}
    \centering
    \includegraphics[width=1\linewidth]{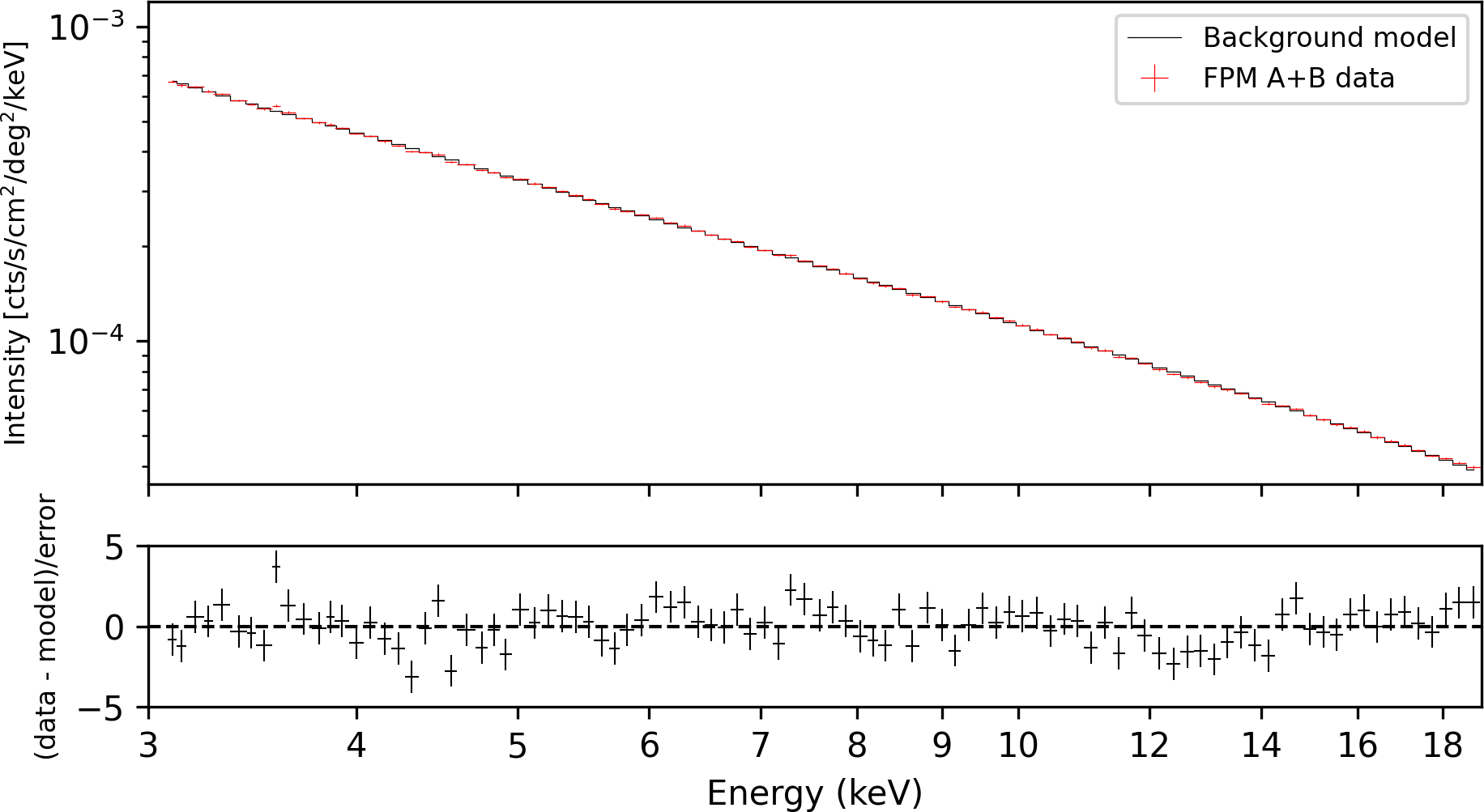}
    \caption{The stacked spectrum of FPMA and FPMB for SL, mesured at $|b| \geq 3^\circ$.}
    \label{fig:raw_spec}
\end{figure}

Then, maximizing the likelihood function of the form:
\begin{equation}
    L = -2\sum_i N_i\log N_{\text{pix},i} - N_{\text{pix},i} - \log N_{i}!
\end{equation}
by the parameters $C_\text{int}$ and $C_\text{apt}$ we find the maximum likelihood estimates for these parameters. Here $N_i$ is the observed number of counts in the $i$-th pixel. The $i$ parameter runs over the array of pixels (excluding bad pixels) for all observations. The final SL spectrum is presented on Fig.\,\ref{fig:raw_spec}.

\subsection{Modeling the Astrophysical Background}
To fit the spectrum presented in Fig. \ref{fig:raw_spec}, we used an XSPEC model of the form \texttt{powerlaw*cflux(highecut*powerlaw)}. The first term gives the long-term averaged contribution from the  solar activity:
\begin{equation}
    I_\text{sol} = N_\text{sol} \left( \frac{E_\gamma}{1\text{ keV}} \right)^{-\Gamma_\text{sol}}.
\end{equation}
Here $N_\text{sol}$ is the normalization (free parameter) and $\Gamma_\text{sol} = 4$ is the photon index of the power law (frozen parameter). The second term gives the contribution of CXB in the 3-20 keV band:
\begin{equation}
    I_\text{CXB} = N_\text{CXB} \left( \frac{E_\gamma}{1\text{ keV}} \right)^{-\Gamma_\text{CXB}} \exp\left(\frac{E^\text{CXB}_\text{cut} - E_\gamma}{E_\text{fold}}\right)
\end{equation}
Here, following the \cite{Gruber:1999yr}, we have fixed a normalization equal to $N_\text{CXB} = 2.4\times10^{-3}$ and photon index $\Gamma_\text{CXB} = 1.29$. We set $E^\text{CXB}_\text{cut} = 10^{-4}$ keV and choose $E_\text{fold}=41.13$ keV as a trial value. The spectrum intensity and model normalization are expressed in units of cts/cm$^2$/s/keV/deg$^2$. While fitting, similar to \cite{Krivonos:2024yvm}, we ignore energy bins above 19 keV due to increased aperture flux caused by the so called Absorbed Stray Light (ASL) \cite{2017JATIS...3d4003M,Rossland:2023vfc,Weng:2024rnt}. Also, systematic uncertainties of the background model were not included in the fit and are therefore not reflected in the quoted parameter errors. The fit is characterised by $\chi^2_r/\text{d.o.f} = 130.0/94 = 1.38$. The high-energy cutoff was estimated at $E^\text{CXB}_\text{cut} = 34.9 \pm 0.6$ keV (hereafter, the uncertainty of each fitting parameter is given for 90\% confidence level). The CXB normalization was measured as $F_{3-20\text{ keV}} = (3.023\pm0.006)\times 10^{-11}$ erg/s/cm$^2$/deg$^2$, that is somewhat higher than the measurements \cite{Krivonos:2020qvl} \footnote{Note that in \cite{Krivonos:2020qvl}, in contrast to the current work and \cite{Krivonos:2024yvm}, we use a significantly different dataset}, but still is consistent with them, taking overall systematic uncertainty into account. Table\,\ref{tab:background_fit} provides a complete summary of our background model.

\begin{table}
    \centering
    \begin{tabular}{llll}
         \hline\hline
         Model & Parameter & Value & Frozen \\
         \hline\hline
         \texttt{powerlaw} & $\Gamma_\text{sol}$ & 4 & True\\
         \texttt{powerlaw} & $N_\text{sol}$ & $(9.8\pm0.3)\times10^{-3}$\footnote{in units of cts/keV/s/cm$^2$/deg$^2$ at 1 keV} & False\\
         \texttt{cflux} & $E_\text{min}$ & 3 keV & True\\
         \texttt{cflux} & $E_\text{max}$ & 20 keV  & True\\
         \texttt{cflux} & Flux & $(3.023 \pm 0.006)$\footnote{in units of $10^{-11}$ erg/s/cm$^2$/deg$^2$} & False\\
         \texttt{powerlaw} & $\Gamma_\text{CXB}$ & 1.29 & True\\
         \texttt{highecut} & $E^\text{CXB}_\text{cut}$ & $10^{-4}$ keV & True\\
         \texttt{highecut} & $E_\text{fold}$ & $34.9 \pm 0.6$ keV & False\\
         \hline\hline
    \end{tabular}
    \begin{tabular}{l}
        Test statistic: $\chi^2_r/\text{d.o.f.}=1.38$, $p=8.27\times10^{-3}$\\
        \hline\hline
    \end{tabular}
    \caption{Spectral parameters, obtained by fitting the background model. All uncertainties are quoted at the 90\% confidence level.}
    \label{tab:background_fit}
\end{table}

\section{Expected Signal from the Galactic DM Annihilation}
The intensity of photons from annihilating DM reads 
\begin{equation}
    I_\gamma = \frac{1}{4\pi}\frac{\langle \sigma v \rangle}{2}\frac{1}{m^2_\chi}\frac{dN}{dE_\gamma}\Biggl \langle \frac{dJ}{d\Omega} \Biggr \rangle.
    \label{intensity}
\end{equation}
Here $\langle \sigma v \rangle$ is the velocity-independent product of the relative velocity $v$ of annihilating particles and $s$-wave cross section $\sigma$, and $m_{\text DM}$ is the mass of DM particle.  The spectrum of photons,  originated in the DM annihilation, $dN/dE_\gamma$, can be approximated for the $2\to 2$ process \eqref{annihilation} as $dN/dE_\gamma=2\delta(E_\gamma - m_\chi)$). The weighted with observation time differential J-factor can be calculated as follows,
\begin{equation}
    \Biggl \langle \frac{dJ}{d\Omega} \Biggr \rangle = \frac{1}{T_\text{tot}}\sum_i T_i\frac{dJ_i}{d\Omega}.
\end{equation}
Here $T_i$ is the exposure time in the $i$-th observation, and the total exposure time is just the sum $T_{\text{tot}} = \sum_i T_i$.

In turn, the differential J-factor for each observation is computed as follows:
\begin{equation}
    \frac{dJ_i}{d\Omega} = \int \rho^2_\text{DM}(r(l_i, b_i, s))ds,
\end{equation}
where $\rho(r)$ refers to the DM radial density profile (we adopt the spherically symmetric halo for the MW), $l_i$ and $b_i$ are galactic latitude and longitude of the $i$-th observation and $s$ denotes the distance from the observer (that is the Earth position) to the point of annihilation. The result of integration over $s$ depends very mildly on the upper limit if it is about the MW virial radius, so we integrate over $s$ in the range $[0, 200]$ kpc. The radial  variable $r$ is expressed in terms of $l$, $b$, and $s$ as:
\begin{equation}
    r(l,b,s)=\sqrt{s^2 + R^2 + 2Rs\cos(l)\cos(b)}, 
\end{equation}
where $R=8.5$ kpc is the distance from the observer to the center of the Milky Way.

\begin{figure}
    \centering
    \includegraphics[width=1\linewidth]{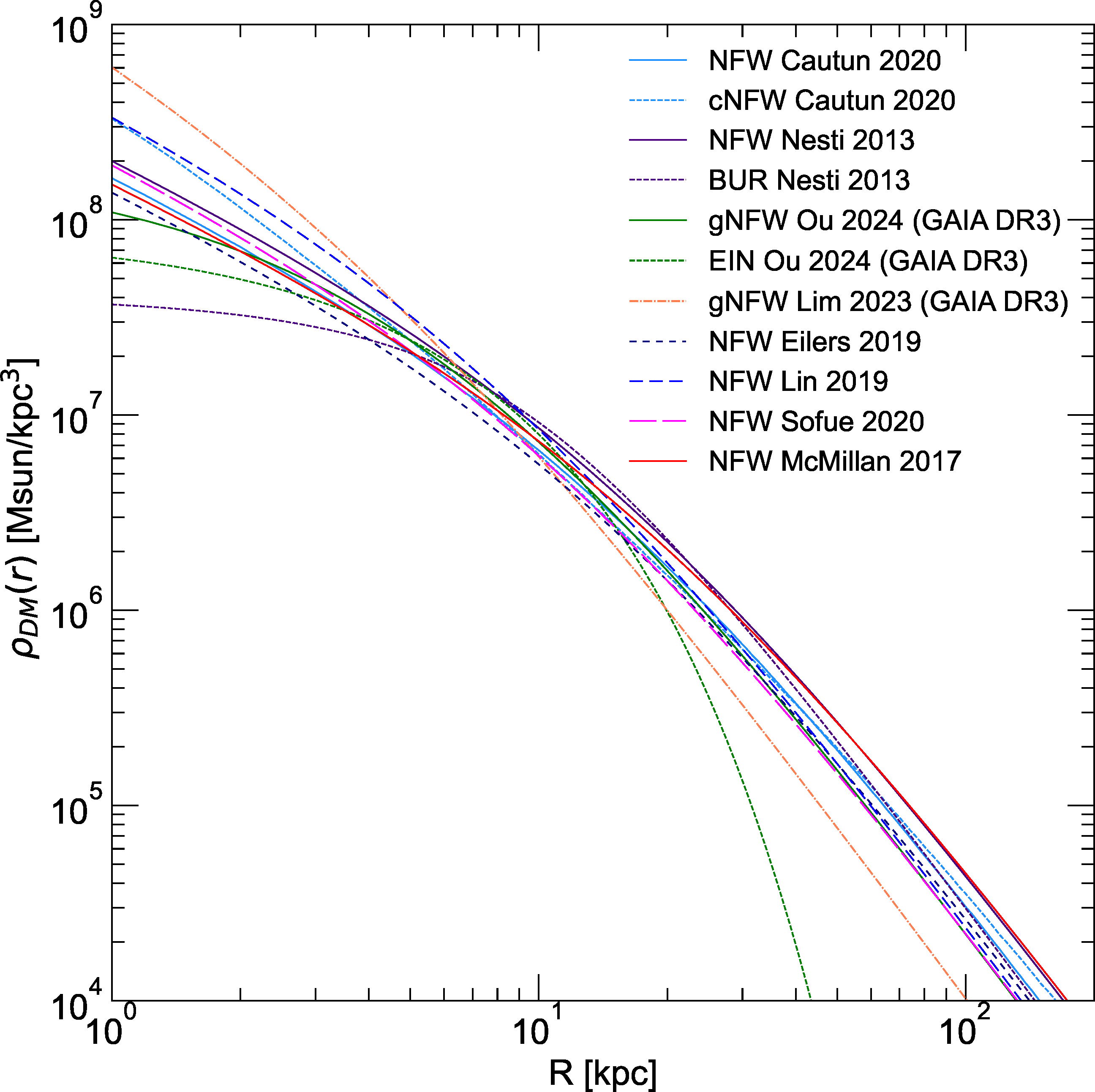}
    \caption{The dark matter profiles we use in our analysis. The base profile is shown in red solid line \cite{McMillan2017}. The remaining profiles are taken from \cite{Eilers2019, Sofue2020, Ou2024, Cautun2020, Lim2023, Nesti2013, Lin2019}.}
    \label{fig:DM_profiles}
\end{figure}

As the basic DM profile of the MW $\rho_{\text DM}(r)$ we chose the standard Navarro--Frank--White (NFW) profile $\rho(r) = \rho_s/(r/r_s)(1+r/r_s)^2$ \cite{NFW} with $\rho_s=8.54\times10^{-3}$ M$_\odot$ pc$^{-3}$ and $r_s=19.6$ kpc \cite{McMillan2017}. However, we also consider other profiles advocated in Refs.\,\cite{Eilers2019, Sofue2020, Ou2024, Cautun2020, Lim2023, Nesti2013, Lin2019} and presented in Fig.\,\ref{fig:DM_profiles}.

\section{Statistical Analysis and Upper Limits}
As it was stated earlier, as a result of the DM annihilation we expect to obtain monochromatic photons and, hence, the photon spectrum can be described by a delta function. However, the motion of DM particles in the Galaxy with characteristic velocities $v \sim 200 - 300$ km/s leads to the Doppler broadening of the line by the value of $\Delta E/E \sim v/c \sim 0.001$. Therefore, we will describe the signal from annihilation using a Gaussian with width $\sigma_{\text{Gauss}} = 0.001$. The DM particles may move slower near the Galaxy center, but it can be safely neglected in our analysis, since even the chosen value of broadening is much smaller than the energy resolution of NuSTAR.  

\begin{figure}
    \centering
    \includegraphics[width=1\linewidth]{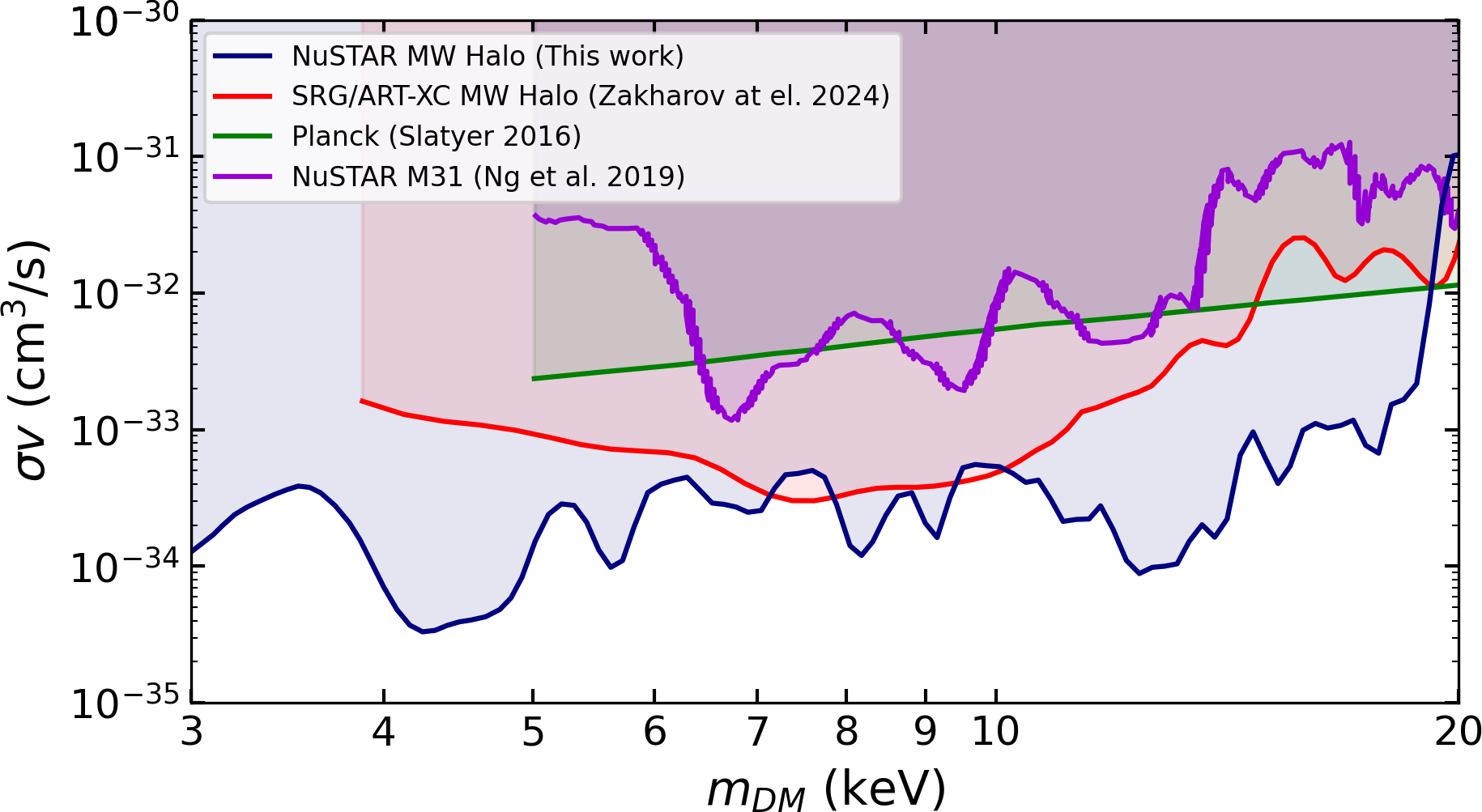}
    \caption{Upper limits (dark blue solid line) on s-wave annihilation cross section (95\% C.L.) obtained with NuSTAR observatory after 11 years of observation the MW halo (for NFW profile \cite{McMillan2017}). The red solid line shows the constraints obtained with SRG/ART-XC after 2 years of operation in survey mode \cite{ART_annihilation}. The dark violet solid line show limits obtained with the NuSTAR in the M31 \cite{nustarM31}. The green solid line show limits obtained with Planck data \cite{Planck}.}
    \label{fig:final_constraints}
\end{figure}

The further procedure of signal search is as follows. To fit the spectrum, we use the XSPEC model in the form of \texttt{powerlaw*cflux(highecut*powerlaw) + gauss}. In this case we assume the width of the Gaussian to be equal to $\sigma_\text{Gauss}$, the position of the Gaussian we consistently change from 3 to 20 keV, and the line intensity of the Gaussian remains the only free parameter that we fit. The minimization procedure yields the intensity of the emission line and the upper limit on the intensity. A one-side upper limit at 95\% confidence level (C.L.) with one degree of freedom implies $\Delta \chi^2 = 2.71$. Thus, we obtain an upper limit on the line intensity. Using Eq. \ref{intensity}, this upper limit can be translated into a constraint on the annihilation cross section $\langle \sigma v \rangle$. The resulting bound on the cross section is shown in Fig. \ref{fig:final_constraints}.

\begin{figure}
    \centering
    \includegraphics[width=1\linewidth]{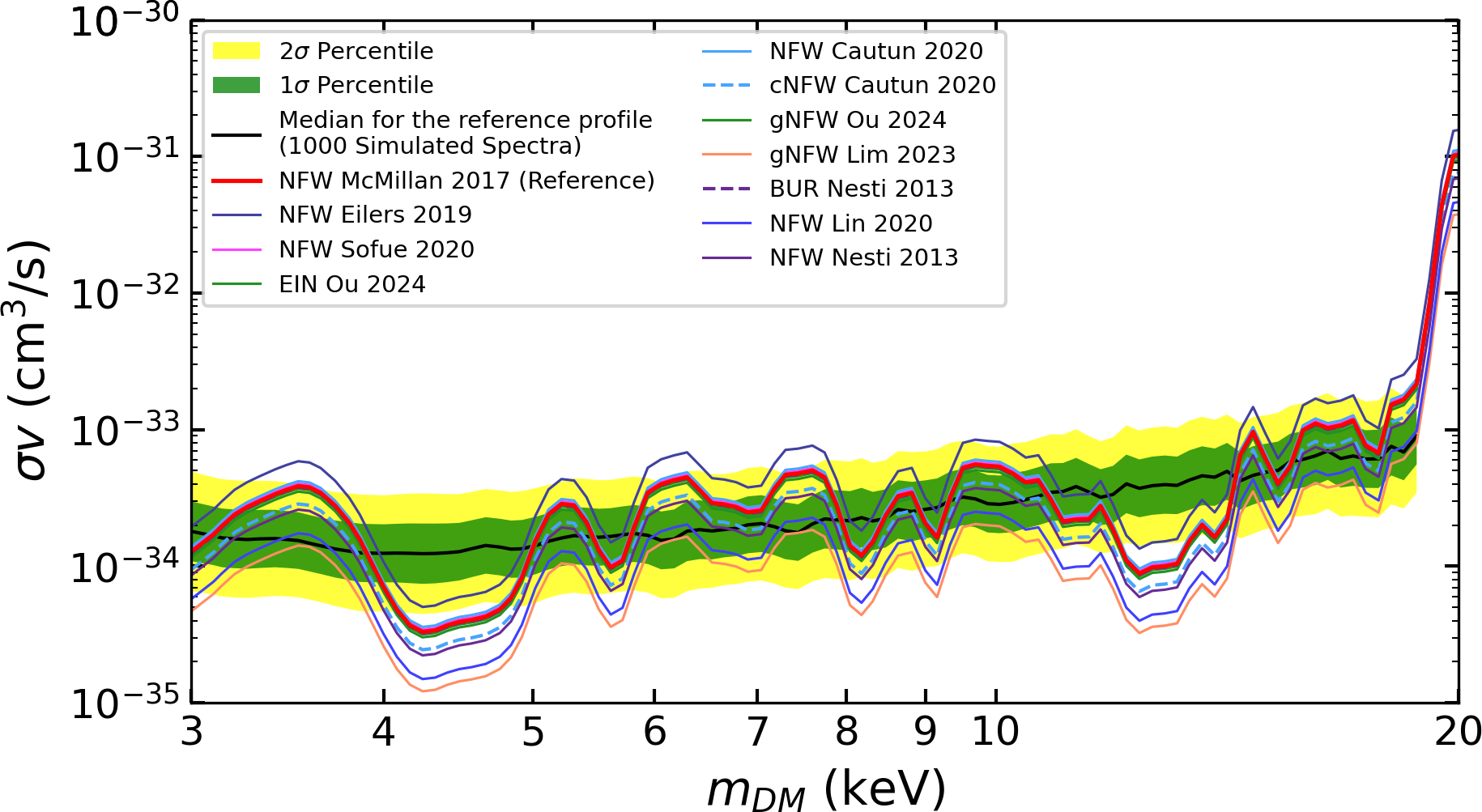}
    \caption{Upper limits on s-wave annihilation cross section (95\% C.L.) obtained with NuSTAR observatory after 11 years of observation the MW halo. The colored lines show the constraints obtained for different profiles \cite{Eilers2019, Sofue2020, Ou2024, Cautun2020, Lim2023, Nesti2013, Lin2019} of the Galactic halo (our reference profile \cite{McMillan2017} is shown as a solid red line). Green and yellow shaded areas show 1$\sigma$ and 2$\sigma$ percentile. The black solid line shows the median value for the expected limit obtained with analysis of 1000 artificial background spectra for reference profile.}
    \label{fig:constraints}
\end{figure}

To obtain constraints on $\langle \sigma v \rangle$, we use the observed X-ray spectrum. Because real data always contain random fluctuations, the shape of the exclusion curve reflects one particular statistical realization. For this reason, it is useful to distinguish between the observed limits, obtained directly from the data, and the expected limits, which show the average sensitivity of the data set. To estimate the expected sensitivity, we performed simulations based on the best-fit spectral model. Using the \texttt{fakeit} tool in XSPEC, we produced $10^3$ artificial spectra and analyzed each of them with the same line-search procedure as applied to the real data. This gave us a distribution of upper limits for each trial line energy. From these ensembles we determined the median expected limit showed as black solid line and the 68\% and 95\% confidence intervals, which are shown as the green and yellow bands in Fig.\ref{fig:constraints}. When comparing the real constraints to these expectations, one finds noticeable deviations in certain mass ranges (e.g. $m_\text{DM} \sim 4-5$ keV and $12-13$ keV). These regions correspond to downward statistical fluctuations in the observed spectrum relative to the smooth background model. Because such dips reduce the available space for an additional line component, the corresponding upper limits appear artificially more stringent. We attribute this effect to moderate systematics in the measured data that are not captured in the simulations.

\section{Conclusion}
Based on the same method and data set presented in our previous work \,\cite{Krivonos:2024yvm}, we have carried out a dedicated search for X-ray signatures of annihilating dark matter in the keV mass range using eleven years of NuSTAR observations in the stray light mode. This observational mode, while originally considered a nuisance for pointed observations, provides a unique opportunity for studies of diffuse emission thanks to its large grasp and stable instrumental background. By carefully constructing a dataset of more than 5200 individual observations with a cumulative usable exposure of 234 Ms, and applying rigorous temporal and spatial filtering, we obtained a broad coverage at intermediate and high Galactic latitudes suitable for precision spectral analysis.

We modeled the combined X-ray spectrum with two background components, the cosmic X-ray background and a solar-scattered contribution. On top of this model we carried out a systematic search for narrow line-like signals in the 3-20 keV range, which would correspond to dark matter annihilation into two photons. No statistically significant excess was detected. As a result, we derived upper limits on the velocity-independent annihilation cross section $\langle \sigma v \rangle$ as a function of the dark matter mass.

Our limits are currently the most stringent constraints in this mass range. They are in good agreement with bounds from other X-ray instruments and further reduce the parameter space available for light annihilating dark matter models. In particular, for typical Galactic halo profiles we exclude cross sections above the level of $\sim 5\cdot10^{-33}$\,cm$^3$\,s$^{-1}$. 

This study highlights the potential of NuSTAR stray light data for indirect searches of dark matter. At the same time, systematic effects such as residual variations of the instrumental background and the exact shape of the cosmic X-ray background remain important sources of uncertainty. Future X-ray missions with higher energy resolution and larger collecting area will provide the sensitivity needed to explore this dark matter window in much greater detail.

\section{Acknowledgements}

D.G. acknowledges the partial support of the work of the National Center of Physics and Mathematics, direction No.5. 

\addcontentsline{toc}{chapter}{\bibname}
\bibliographystyle{apsrev4-1}
\bibliography{bibliography}

\end{document}